\documentclass[%
 reprint,
 amsmath,amssymb,
 aps,
]{revtex4-2}

\usepackage{graphicx}
\usepackage{dcolumn}
\usepackage{bm}
\usepackage{subcaption}
\usepackage{hyperref}
\usepackage{verbatim}   
\usepackage{physics}    
\usepackage{cleveref}   
\usepackage{xcolor}
\usepackage{hyperref}   

\begin{document}

\preprint{APS/123-QED}

\title{Exact demagnetisation field for periodic one-dimensional array of rectangular prisms}

\author{Frederik Laust Durhuus}
\affiliation{
 Department of Energy Conversion and Storage, Technical University of Denmark - DTU, DK-2800 Kgs. Lyngby, Denmark
}%

\author{Andrea Roberto Insinga}
\affiliation{
 Department of Energy Conversion and Storage, Technical University of Denmark - DTU, DK-2800 Kgs. Lyngby, Denmark
}%


\author{Rasmus Bjørk}
 \email{rabj@dtu.dk}
\affiliation{
 Department of Energy Conversion and Storage, Technical University of Denmark - DTU, DK-2800 Kgs. Lyngby, Denmark
}%

\date{\today}

\begin{abstract}
The magnetic field from a uniformly magnetised, rectangular prism is known exactly, which is the basis for a large number of micromagnetic simulations. Here we derive an analytical solution for the field from a periodically repeating infinite array of prisms aligned end-to-end, which becomes exact on the center axis in the limit of infinitesimally thin prisms. Using the same method we derive the on-axis field for a one-dimensional array of point dipoles. We validate the obtained results numerically and furthermore compare with the common macrogeometry approach and more recent uniform magnetisation method, demonstrating an excellent convergence rate for the novel method. 
\end{abstract}

\maketitle

\section{Introduction}

One of the most challenging aspects in modelling magnetic materials are the magnetostatic interactions, i.e.\ the long-range, dipolar interaction between each volume element of magnetised material \cite{Abert_2013}.

This is particularly true in micromagnetic computations, where simulating macroscopic samples is often computationally infeasible because of the number of interactions within the modelled system. The fundamental assumption in micromagnetism is that the system can be discretized into cells with uniform magnetization.
This is valid in sufficiently small cells, typically nm-scale, where ferromagnetic exchange coupling dominates, ensuring aligned spins within each cell\cite{Donahue_1999,Vansteenkiste_2014,bjork_magtense_2021}. The assumption also holds for sufficiently small ferromagnetic particles which are essentially single-domain\cite{aharoni_elongated_1988,aharoni_single-domain_1989,brown_fundamental_1968,brown_fundamental_1969}, with the shape anisotropy from magnetostatic self-interaction being an essential contribution to their dynamics and field response\cite{durhuus_magnetic_2024}. 

To alleviate the problem of modelling magnetic interactions across a whole sample, periodic boundary conditions (PBCs) can be employed, where displaced copies of the simulated domain are used to approximate the remaining sample\cite{lebecki_periodic_2008}. A common implementation is the macrogeometry approach\cite{fangohr_new_2009} as employed e.g.\ in mumax3 \cite{vansteenkiste_design_2014}, where a finite number of domain copies are used in a one-off calculation before the micromagnetic simulation to modify the field calculation through the demagnetization tensor. Meanwhile, in the OOMMF software, analytical solutions for distant domain copies are used to implement PBCs extending to infinity in 1D\cite{lebecki_periodic_2008} and 2D\cite{wang_two-dimensional_2010}, however these rely on a combined dipolar- and continuum approximation.

In this paper we consider an infinite one-dimensional (1D) array of identical rectangular prisms, aligned end-to-end along $x$ as illustrated in \cref{fig:thin_prism_array}, and derive a closed-form, analytical expression for the magnetic field which is exact on the $x$-axis in the limit of infinitesimally thin prisms. With the same techniques, we also solve the on-axis field for a 1D array of point dipoles. 

We demonstrate numerically that our prism solution is an excellent approximation to the field from the infinity of distant domain copies and when used to supplement the macrogeometry approach, the number of copies in the macrogeometry required for convergence is greatly reduced. Finally, we compare to the recently developed uniform magnetisation approach, where the magnetisation of distant domain copies are approximated by their average\cite{durhuus_shape_2026}, which is also applicable to 2D and 3D PBCs. While both improve on the base macrogeometry method, the convergence rate is superior for the analytical sum of prisms, especially when the simulation domain is much larger in the direction of the PBC compared to the transverse directions.

\section{Theory}

\begin{figure*}[htb]
    \centering
    \includegraphics[width=\linewidth]{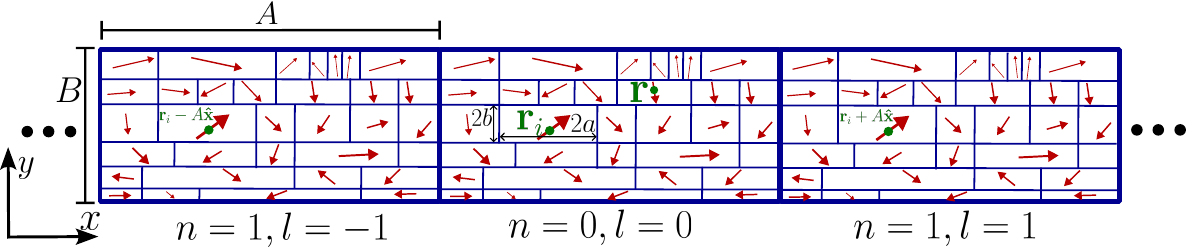}
    \caption{2D sketch of periodic boundary conditions along the $x$-axis as in e.g. a micromagnetic simulation. The initial domain ($n=0$) and the nearest pair of copies ($n=1$) with dots indicating that the system is repeated ad infinitum along $x$. The variable $l$ is the domain index, which is positive for positive $x$ and vice versa. Arrows represent magnetisation vectors for each prism-shaped micromagnetic cell. The goal in this study is to evaluate the magnetic field at a given point $\vb{r}$ from a given cell centered at $\vb{r}_i$ and its infinite copies $\vb{r}_i \pm lA\vb{\hat{x}}$ (all marked with green dots).}
    \label{fig:1D_PBC_illustrated}
\end{figure*}

\subsection{The micromagnetic stray field problem with periodic boundary conditions \label{subsec:micromagnetics_problem}}

Given a collection of micromagnetic cells, which by definition are uniformly magnetised, the magnetic field $\vb{H}$ at position $\vb{r}$ is given by
\begin{align}
    \vb{H}(\vb{r}) = - \sum_{i} \mathrm{N}_{i}(\vb{r} - \vb{r}_{i}) \vb{M}(\vb{r}_{i}), \label{eq:H_field}
\end{align}
where $i$ is cell index, $\vb{M}$ is magnetisation, $\vb{r}_{i}$ is the center position of cell $i$ and $\mathrm{N}_{i}$ is the point demagnetisation tensor which is a dimensionless, symmetric 3-by-3 tensor that exclusively depends on the shape of cell $i$.

The point demagnetization tensor is known for the triaxial ellipsoids\cite{stoner_xcvii_1945,osborn_demagnetizing_1945}, which has a uniform internal field\cite{maxwell_treatise_1891}, but also for e.g. uniformly magnetised cylinders\cite{caciagli_exact_2018}, rectangular prisms\cite{smith_demagnetizing_2010}, triaxial ellipsoids\cite{tejedor_external_1995}, tetrahedra\cite{nielsen_stray_2019} and hemispheres\cite{durhuus_demagnetization_2025}, with the hemisphere results recently used to model ferrofluid droplets\cite{akbari_impact_2025}. At distances far away from each of these geometries, the field approaches that of a dipolar field \cite{Bjork_2023}.

We wish to solve the problem of infinite, periodic boundary conditions (PBCs) along the $x$-axis, which means the whole simulated domain is copied at intervals of say $A$ along the entire $x$-axis as illustrated in \cref{fig:1D_PBC_illustrated} and these copies are included in the summation in \cref{eq:H_field}. Note that the initial domain can include a vacuum layer or other non-magnetic regions to model e.g.\ porous materials or arrays of discrete magnets. 

Let $l$ be the domain index, and $i$ the internal cell index for the $l=0$ domain, so by the PBC, $\vb{M}(\vb{r}_i + l A \vb{\hat{x}}) = \vb{M}(\vb{r}_i)$, hence
\begin{align}
    \vb{H}(\vb{r}) = - \sum_{i} \mathrm{N}_{i}^\text{PBC}(\vb{r} - \vb{r}_{i}) \vb{M}(\vb{r}_{i}),  \label{eq:H_PBC}
\end{align}
where the domain copies are encoded in
\begin{align}
    \mathrm{N}_{i}^\text{PBC}(\vb{r} - \vb{r}_{i}) = \sum_{l=-\infty}^\infty \mathrm{N}_{i}(\vb{r} - \vb{r}_{i} - lA\vb{\hat{x}}). \label{eq:N_PBC}
\end{align}
Thus the problem of introducing 1D-PBCs is reduced to performing the sum over $l$ from \cref{eq:N_PBC} for each cell once, then the modified demagnetisation tensors can be stored for subsequent micromagnetic computations. However, the infinite sum can rarely be done analytically, so we define the remainder $\mathrm{N}^\text{rem}_i$ by
\begin{align}
     \mathrm{N}_{i}^\text{PBC} (\vb{r} - \vb{r}_{i}) = \sum_{l=-n}^n \mathrm{N}_{i}(\Delta \vb{r}_l) + \mathrm{N}^\text{rem}_{i}(\vb{r} - \vb{r}_{i}),  \label{eq:N_rem}
\end{align}
where
\begin{align}
    \Delta \vb{r}_l = \vb{r} - \vb{r}_i - l A \vb{\hat{x}}. \label{eq:Delta_r}
\end{align}

The sum in \cref{eq:N_rem} can numerically relatively easily be performed for the $2n+1$ most central copies, i.e.\ the initial domain and $n$ surrounding copies. Then the remaining infinity of copies are all far away along the $x$-axis, which justifies several analytically solvable approximations to $\mathrm{N}^\text{rem}$ 
which we discuss in the following. 


\subsection{Array of point dipoles \label{subsec:array_of_point_dipoles}}

\begin{figure}[h]
    \centering
    \includegraphics[width=\linewidth]{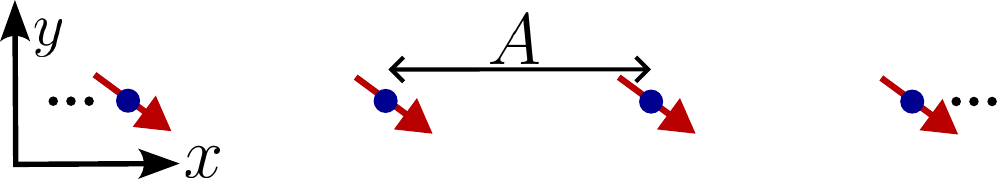}
    \caption{Illustration of an array of identical point dipoles placed at intervals of $A$ along the $x$-axis.}
    \label{fig:point_dipole_array}
\end{figure}

We start with the simplest periodic system, i.e.\ an array of point dipoles placed on the $x$-axis at intervals of $A$ and each having the same magnetic moment $\vb{m}$ (see \cref{fig:point_dipole_array}). The magnetic field at position $\vb{r}$ from a dipole at the origin is (see e.g.\ Eq. (5.89) of \cite{griffiths_introduction_2013})
\begin{align}
    \vb{H}_\text{dip} = \frac{1}{4\pi r^3} \left[3 (\vb{\hat{r}} \vdot \vb{m}) \vb{\hat{r}} - \vb{m} \right],
\end{align}
which may be written
\begin{align}
    \vb{H}_\text{dip} = -\mathrm{N}_\text{dip} \vb{m},  \label{eq:H_dip}
\end{align}
where the demagnetisation tensor is
\begin{align}
    \mathrm{N}_\text{dip} = -\frac{1}{4\pi r^5}\mqty(3x^2 - r^2 & 3xy & 3 xz \\ 3xy & 3y^2 - r^2 & 3yz \\ 3xz & 3yz & 3 z^2 - r^2). \label{eq:N_dip}
\end{align}
and $r = \sqrt{x^2+y^2+z^2}$. When restricting the field evaluation to points where $x \gg y, z$ so that $r \approx \abs{x}$ the tensor simplifies to
\begin{align}
    \mathrm{N}_\text{dip} \approx - \frac{1}{4\pi \abs{x}^3} \mqty(2 & 3 \frac{y}{x} & 3 \frac{z}{x} \\ 3 \frac{y}{x} & 3 \frac{y^2}{x^2} - 1 & 3 \frac{yz}{x^2} \\ 3 \frac{z}{x} & 3 \frac{yz}{x^2} & 3 \frac{z^2}{x^2} - 1).  \label{eq:N_dip_simplified}
\end{align}

Now consider a dipole at position $\vb{r}_i$ and assume the magnetic field from itself and its $2n$ closest copies is evaluated exactly using \cref{eq:N_dip}. The contribution from the remainder of the infinite dipole array is described by
\begin{align}
    \mathrm{N}^\text{rem}_\text{dip} &=\sum_{l=-\infty}^{-(n+1)}\mathrm{N}_\text{dip}(\Delta \vb{r}_l) + \sum_{l=n+1}^\infty \mathrm{N}_\text{dip}(\Delta \vb{r}_l),    \label{eq:N_rem_dip}
\end{align}
with $\Delta \vb{r}_l$ defined in \cref{eq:Delta_r}. Thus, using \cref{eq:N_dip_simplified} and noting that $y,z$ are equal for all $l$, we need to sum $\frac{1}{(x - x_i - lA)^q}$ over $l$ for $q \in \{3,4,5\}$. To this end we use the $m$'th order polygamma function, defined by\cite{arfken_mathematical_2005}
\begin{align}
    \Psi_m(x) = \dv[m+1]{x} \ln x! = (-1)^{m+1} m! \sum_{l=0}^\infty \frac{1}{(x + l)^{m+1}}. \label{eq:polygamma}
\end{align}
From the above expression and the recurrence relation
\begin{align*}
    \Psi_m(x+1) = \Psi_m(x) + \frac{(-1)^m m!}{x^{m+1}},
\end{align*}
one may show that for $n=0,1,2,...$
\begin{align}
    &\Psi_m(x + n) = (-1)^{m+1} m! \sum_{l=n}^\infty \frac{1}{(x + l)^{m+1}}, 
\end{align}
hence
\begin{align}
    \sum_{l=n}^\infty \frac{1}{(x \pm lA)^{m+1}} = \frac{(-1)^{m+1} \Psi_m\left(\frac{x}{A} + n\right)}{(\pm A)^{m+1} m!}. \label{eq:Polygamma_identity}
\end{align}
Applying this result to \cref{eq:N_dip_simplified,eq:N_rem_dip}, it follows that
\begin{align}
    \mathrm{N}^\text{rem}_\text{dip} \approx \frac{1}{4\pi A^3} \mqty(\psi_2 & \frac{y}{2A}\psi_3 & \frac{z}{2A}\psi_3 \\ \frac{y}{2A}\psi_3 & \frac{y^2}{8 A^2}\psi_4 - \frac{1}{2} \psi_2 & \frac{yz}{8A^2} \psi_4 \\ \frac{z}{2A} \psi_3 & \frac{yz}{8A^2} \psi_4 & \frac{z^2}{8 A^2}\psi_4 - \frac{1}{2} \psi_2), \label{eq:N_rem_dip_solution}
\end{align}
where we define the function $\psi_m$ as
\begin{align}
    \psi_m(\vb{r} - \vb{r}_i, n) = \sum_{p=\pm 1} \Psi_m\left(p^n\frac{x - x_i}{A} + n + 1\right).  \label{eq:psi_m}
\end{align}
Since the stray-field of any micromagnetic cell is nearly dipolar at long range \cref{eq:H_PBC,eq:N_dip,eq:N_rem,eq:Delta_r,eq:N_rem_dip_solution,eq:psi_m} with $\vb{M}_i \rightarrow \vb{m}_i$ constitute a full, approximate solution for the micromagnetic stray-field with 1D-PBCs (cf.\ \cref{fig:1D_PBC_illustrated}) where $n$ is a free parameter to converge numerically and $\vb{m}_i = V_i \vb{M}_i$ with $V_i$ denoting the volume of cell $i$. 

For the simpler case of interactions within a single dipole array (\cref{fig:point_dipole_array}) we can set $y,z=0$ and $n=0$ without loss of generality. Thus the exact field from a dipole at position $x_i$ and its infinite copies, evaluated at $x$, is
\begin{align}
    \vb{H}_\text{dip}^\text{array}(x) = \frac{1}{4\pi}\left(\frac{1}{\abs{x - x_i}^3} - \frac{\psi_2(x - x_i,0)}{A^3}\right) \mathrm{K} \vb{m}, 
\end{align}
where we defined the 3-by-3 matrix
\begin{align}
    \mathrm{K} = \mqty(2 & 0 & 0 \\ 0 & -1 & 0 \\ 0 & 0 & -1).  \label{eq:K_matrix}
\end{align}
The first term in parentheses is the dipoles own field (cf. \cref{eq:H_dip,eq:N_dip}) and the second a correction from its copies. Of particular interest is the self-interaction in the array, i.e.\ the field on moment $i$ from its copies:
\begin{align}
    \vb{H}^\text{int}_\text{dip} = -\frac{\psi_2(0,0)}{4\pi A^3} \mathrm{K} \vb{m} = \frac{\zeta(3)}{\pi A^3} \mathrm{K} \vb{m},
\end{align}
where $\zeta$ is the Riemann-Zeta function and $\zeta(3) \approx 1.2021$. Thus the interaction energy per dipole is
\begin{align}
    \vb{U}_\text{dip}^\text{int} = - \vb{H}_\text{dip}^\text{int} \vdot \vb{m} = \frac{m^2 \zeta(3)}{4\pi A^3} (1 - 3 \cos^2 \theta),   \label{eq:U_dip_int}
\end{align}
where $\theta$ is the angle of $\vb{m}$ wrt.\ the $x$-axis. \Cref{eq:U_dip_int} describes an interaction-induced anisotropy - analogous to the shape anisotropy of 3D magnets - which favours the magnetisation pointing along the axis of the dipole array.

\subsection{Array of rectangular prisms \label{subsec:array_of_prisms}}

\begin{figure}[h]
    \centering
    \includegraphics[width=\linewidth]{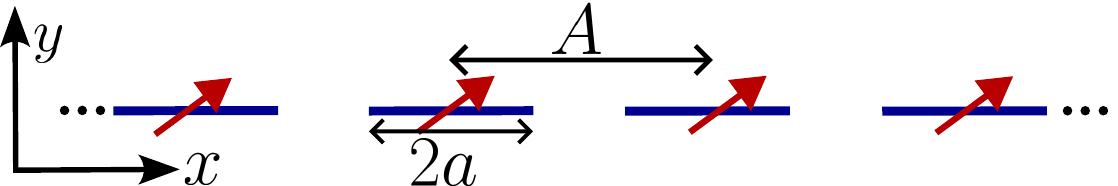}
    \caption{Illustration of special case where the analytical solution from \cref{subsec:array_of_prisms} becomes exact, i.e.\ when calculating the internal field for an array of identical, thin rectangular prisms end-to-end aligned and placed at intervals of $A$ along the $x$-axis.}
    \label{fig:thin_prism_array}
\end{figure}

We now consider an array of rectangular prisms as in \cref{fig:thin_prism_array}. For a single such prism centered on the origin, the diagonal components of the demagnetisation tensor are given by\cite{smith_demagnetizing_2010}
\begin{align}
    N_{\alpha \alpha}^\text{prism}(\vb{r}) = \frac{1}{4\pi} \sum_{p_x,p_y,p_z} \arctan f_\alpha (p_x x, p_y y, p_z z),  \label{eq:N_prism_diagonal}
\end{align}
where $p_x, p_y, p_z = \pm 1$ and we sum over all 8 combinations, while
\begin{align}
    f_\alpha(\vb{r}) = \frac{\prod_{\beta \neq \alpha} (a_\beta - r_\beta)}{(a_\alpha - r_\alpha) \abs{\vb{a} - \vb{r}}},
\end{align}
with $\alpha, \beta \in \{x,y,z\}$ indexing Cartesian components and $\vb{a} = (a,b,c)$ being half the prism side-lengths along $x,y$ and $z$ as illustrated in \cref{fig:1D_PBC_illustrated}. For the off diagonal components ($\alpha \neq \beta$)\cite{smith_demagnetizing_2010}
\begin{align}
    N_{\alpha \beta}^\text{prism} = \sum_{p_x,p_y,p_z} \frac{-p_x p_y p_z}{4\pi} \ln F_{\alpha \beta}(\vb{r}, p_x a, p_y b, p_z c),    \label{eq:N_prism_off_diagonal}
\end{align}
where
\begin{align}
    F_{\alpha \beta}(\vb{r}, a, b, c) = a_\gamma - r_\gamma + \abs{\vb{a} - \vb{r}},   \label{eq:F_alpha_beta}
\end{align}
and $\gamma \in \{x,y,z\}$ is different from $\alpha, \beta$, for example if $\alpha=x$ and $\beta = y$ then $\gamma = z$.

Without loss of generality we write 
\begin{align}
    \abs{\vb{a} - \vb{r}} = \abs{a - x} \sqrt{1 + \epsilon}, \quad \epsilon = \frac{(b - y)^2 + (c - z)^2}{(a - x)^2}.    \label{eq:small_number}
\end{align}
The core simplification which makes \cref{eq:N_prism_diagonal,eq:N_prism_off_diagonal} analytically summable is that 
\begin{align}
    \abs{a - p_x x} \gg \abs{b - p_y y}, \abs{c - p_z z},   \label{eq:core_simplification}
\end{align}
which implies that $\epsilon \ll 1$ and thus justifies Taylor expanding $\mathrm{N}^\text{prism}$ to leading order in $\epsilon$.

The self-interaction can be handled exactly by \cref{eq:N_prism_diagonal,eq:N_prism_off_diagonal}, so we can assume $\abs{x} > a$ when developing the simplified demagnetisation tensor. It follows that
\begin{align}
    \abs{a - p_x x} = - p_x \mathrm{sign}(x) (a - p_x x)   \label{eq:sign_relation}
\end{align}
where the $\mathrm{sign}$-function returns the sign of its argument.

Applying \cref{eq:core_simplification,eq:small_number,eq:sign_relation} to the $yy$ component of \cref{eq:N_prism_diagonal} we find that the zeroth order term in $N_{yy}^\text{prism}$
\begin{align*}
    &\frac{1}{4\pi} \sum_{p_x,p_y,p_z} \arctan \frac{(a - p_x x)(c - p_z z)}{(b-p_y y) \abs{a - p_xx}}
    \notag \\
    &= - \frac{\mathrm{sign}(x)}{4\pi} \sum_{p_x, p_y, p_z} p_x \arctan \frac{c - p_z z}{b - p_y y}
    \notag \\
    &= 0,
\end{align*}
where the last equality follows from cancellation between the $p_x = +1$ and $p_x = -1$ terms. However, Taylor expanding to first order, we get the leading order approximation
\begin{align}
    N_{yy}^\text{prism}(\vb{r}) &\approx \frac{\mathrm{sign}(x)}{8\pi} \sum_{p_x,p_y,p_z} \frac{(b - p_y y)(c - p_z z) \epsilon}{(b - p_y y)^2 + (c - p_z z)^2}
    \notag\\
    &=\frac{bc\: \mathrm{sign}(x)}{2\pi} \left[\frac{1}{(a - x)^2} - \frac{1}{(a + x)^2} \right],
\end{align}
and because the expression is symmetric in $b,c$ and $y,z$ we have $N_{zz}^\text{prism} \approx N_{yy}^\text{prism}$. As for the $xx$ component, the zeroth order term is non-zero, so
\begin{align*}
    N_{xx}^\text{prism}(\vb{r}) &\approx \frac{1}{4\pi} \sum_{p_x,p_y,p_z} \arctan \frac{(b - p_y x)(c - p_z z)}{(a-p_x y) \abs{a - p_xx}}
    \notag \\
    &\hspace{-1cm} = - \frac{\mathrm{sign}(x)}{4\pi} \sum_{p_x, p_y, p_z} p_x \arctan \frac{(b - p_y y)(c - p_z z)}{(a - p_x x)^2}.
\end{align*}
It follows from \cref{eq:core_simplification} that the argument of the arctan function is $\ll 1$, so we can Taylor expand to leading order using $\arctan x = x + \order{x^3}$, which yields
\begin{align}
    N_{xx}^\text{prism}(\vb{r}) \approx \frac{bc \: \mathrm{sign}(x)}{\pi} \left[\frac{1}{(a + x)^2} - \frac{1}{(a - x)^2}\right].
\end{align}

For the off-diagonal elements \cref{eq:N_prism_off_diagonal,eq:F_alpha_beta}, we have that
\begin{align}
    \ln F_{\alpha \beta} = \eval{\ln F_{\alpha \beta}}_{\epsilon = 0} - \frac{\abs{a - x} \epsilon}{2 (a_\gamma - r_\gamma + \abs{a - x})} + \order{\epsilon^2}.   \label{eq:F_taylor}
\end{align}

We note that
\begin{align*}
    \eval{F_{xy}(\vb{r}, a, b, c)}_{\epsilon = 0} = c - z + \abs{a - x},
\end{align*}
is not a function of $b$ so when summing over $p_y$ the terms cancel pairwise. The same sort of cancellation happens for the other 5 off-diagonal tensor-components because the dependence on $b$ or $c$ is lost when $\epsilon = 0$. The first order term in $N_{xy}^\text{prism}$ is (cf. \cref{eq:N_prism_off_diagonal,eq:F_taylor})
\begin{align*}
    \sum_{p_x,p_y,p_z} &\frac{p_x p_y p_z \abs{p_xa - x}\:  [(p_y b - y)^2 + (p_z c - z)^2]}{8\pi \left(p_y b - y + \abs{p_x a - x} \right) (p_x a - x)^2}
    \\
    &= \frac{1}{2\pi} \sum_{p_x,p_y} \frac{p_x p_y \mathrm{sign}(x) \: c z}{\left(p_y b - y + \abs{p_x a - x} \right)(p_x a - x)} 
    \\
    &= - \frac{1}{\pi} \sum_{p_x} \frac{p_x \mathrm{sign}(x) \: bc z}{(p_x a - x) [(\abs{p_x a - x} - y)^2 - b^2]}.
\end{align*}
Thus, using $\abs{p_x a - x} \gg \abs{y}, b$ to simplify the denominator, we find that to leading order in $\epsilon$
\begin{align*}
    N_{xy}^\text{prism}(\vb{r}) \approx \frac{bc z \: \mathrm{sign}(x)}{\pi} \left[\frac{1}{(x - a)^3} - \frac{1}{(x + a)^3} \right].
\end{align*}
We note that $N_{xy}^\text{prism} \sim \frac{z}{a \pm x} N_{xx}^\text{prism}$ so it is relatively negligible away from the prism end points. By symmetry, the same holds for the $yx, xz$ and $zx$ components while a similar calculation shows that $N_{yz}^\text{prism}, N_{zy}^\text{prism} = \order{\epsilon^2}$. 

In summary, under the assumption of \cref{eq:core_simplification}, the external magnetic field of a uniformly magnetised prism is approximately
\begin{align}
    \mathrm{N}_\text{prism} \approx \mathrm{sign}(x) \frac{bc}{\pi} \left[\frac{1}{(a+x)^2} - \frac{1}{(a-x)^2} \right] \mathrm{K}  \label{eq:N_prism_simplified}
\end{align}
where $\mathrm{K}$ is given in \cref{eq:K_matrix}. Comparing to \cref{eq:N_dip_simplified} we see that to leading order in $\frac{y}{x}, \frac{z}{x}$ the relative values of the tensor components are the same as for a point dipole, which makes sense as the prism field should be nearly dipolar at long range. Also, like the dipole case, the off-diagonal elements are small compared to the diagonal due to a stronger $\frac{1}{\abs{x}}$ dependence.

Now for a prism at position $\vb{r}_i$ and a periodic array of copies displaced at intervals of $A$ along the $x$-axis as in \cref{fig:thin_prism_array}, we can write the approximate demagnetisation tensor for each copy as
\begin{align}
    \mathrm{N}_\text{prism}(\Delta \vb{r}_l) \approx \mathrm{sign}(x - x_i - lA) \frac{bc}{\pi} g_l \mathrm{K},
\end{align}
where $\Delta \vb{r}_l$ is defined in \cref{eq:Delta_r} and we define $g_l$ by
\begin{align}
    g_l = \left[\frac{1}{(a + x - x_i - lA)^2} - \frac{1}{(a - x + x_i + lA)^2} \right].
\end{align}
Applying the polygamma function identity \cref{eq:Polygamma_identity} to perform the sum over $l$ for all copies displaced $nA$ or further from the original, we arrive at the main result of this paper:
\begin{align}
    \mathrm{N}_\text{prism}^\text{rem}(\vb{r} - \vb{r}_i) &\approx \frac{bc}{\pi} \mathrm{K}\left(\sum_{l=n+1}^\infty g_l - \sum_{l=-\infty}^{-n-1}g_l\right) 
    \notag\\
    &\hspace{-1.7cm} = \frac{bc}{\pi A^2} \mathrm{K} \sum_{p_a,p_x = \pm 1} p_a \Psi_1\left(\frac{p_a a + p_x (x - x_i)}{A} + n + 1 \right).    \label{eq:N_rem_prism}
\end{align}

In the calculation of $\mathrm{N}^\text{rem}$ for micromagnetics, where the exact field is used for the $2n+1$ most central prism copies, the core assumption \cref{eq:core_simplification} is valid in the entire simulated domain when $n$ is large enough, which justifies using \cref{eq:N_rem_prism} for the demagnetisation contribution of distant copies. The trigamma function $\Psi_1$ is easily computed with standard software like the Python package scipy\cite{virtanen_scipy_2020}, so \cref{eq:N_rem,eq:N_prism_diagonal,eq:N_prism_off_diagonal,eq:N_rem_prism,eq:H_PBC,eq:Delta_r} constitute a complete and practical solution to the micromagnetic stray field with 1D PBCs.

We emphasize that for a thin prism ($b/a, c/a \approx 0$) \cref{eq:core_simplification} is nearly exact everywhere on the prism center axis ($y - y_i, z - z_i = 0$) except the prism endpoints where $x - x_i = \pm a$. At $a=A/2$ the prism copies merge, so the field from the two nearest copies also diverge at $\Delta x = \pm a$ when using the approximated $\mathrm{N}^\text{prism}$, which is reflected in \cref{eq:N_rem_prism}. Thus, to avoid the endpoint singularities we recommend using the full \cref{eq:N_prism_diagonal,eq:N_prism_off_diagonal} for the prism self-interaction in all cases and for the first pair of copies when $a \approx A/2$. With \cref{eq:N_rem_prism} for the remaining infinity of copies, \cref{eq:H_PBC} yields the exact internal field for an array of infinitesimally thin rectangular prisms of magnetisation $\vb{M}$ as sketched in \cref{fig:thin_prism_array}.

\subsection{Uniform magnetisation approximation \label{subsec:uniform_magnetisation}}

\begin{figure}[h]
    \centering
    \includegraphics[width=\linewidth]{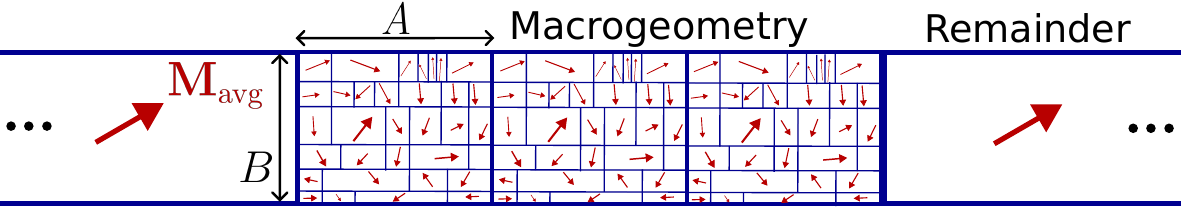}
    \caption{Illustration of uniform magnetisation approximation with $n=1$, i.e.\ the macrogeometry consists of the simulated domain plus 1 domain copy. The remainder of the system is approximated as a pair of uniformly magnetised, semi-infinite rectangular prisms with the same average magnetisation $\vb{M}_\text{avg}$. The domain length along $z$ is denoted $C$.}
    \label{fig:uniform_magnetisation}
\end{figure}

For comparison we here derive an alternative approach to micromagnetic PBCs. Following Ref.\ \cite{durhuus_shape_2026} we use the macrogeometry method for the $2n$ innermost copies and approximate the remainder as a uniformly magnetised continuum with the same average magnetisation $\vb{M}_\text{avg}$ as the initial domain (see \cref{fig:uniform_magnetisation}). We proved in Ref.\ \cite{durhuus_shape_2026} that this method converges to the correct field for both 1D, 2D and 3D PBCs as $n \rightarrow \infty$ even when the initial domain has non-magnetic regions.

In the present case, with 1D PBCs, a rectangular simulation domain and the nearest $n$ domain copies on both sides treated exactly, the remainder is a pair of uniformly magnetised, semi-infinite rectangular prisms with transverse dimensions $B \times C$ (cf. \cref{fig:uniform_magnetisation}). By the linearity of magnetic fields and demagnetisation tensors, the magnetostatic field at $\vb{r}$ thus becomes
\begin{align}
    \vb{H}(\vb{r}) \approx &-\sum_{i} \sum_{l=-n}^n \mathrm{N}_{i}(\vb{r} - \vb{r}_{i} - lA\vb{\hat{x}}) \vb{M}_{i} 
    \notag\\
    &- [\mathrm{N}_\infty^\text{prism}(\vb{r}) - \mathrm{N}_n^\text{prism}(\vb{r})] \vb{M}_\text{avg},
\end{align}
where $\mathrm{N}_\infty^\text{prism}(\vb{r})$ is the demagnetisation tensor for the entire system, i.e.\ a rectangular prism infinite along $x$, while $\mathrm{N}_n^\text{prism}$ is the tensor for the macrogeometry i.e.\ the $2n+1$ most central domain copies. 

$\mathrm{N}_n^\text{prism}$ is found from \cref{eq:N_prism_diagonal,eq:N_prism_off_diagonal} by letting $[a,b,c] \rightarrow [(2n+1)A, B, C]$. To calculate $\mathrm{N}_\infty^\text{prism}$, we note that when $a \rightarrow \infty$ and $b,c \rightarrow B,C$ in \cref{eq:N_prism_diagonal,eq:N_prism_off_diagonal}, the result is $N_{\alpha\beta} = 0$ when $\alpha\neq \beta$, while for the diagonal elements
\begin{align}
    f_x \rightarrow \lim_{a \rightarrow \infty} \frac{(B-y)(C-z)}{a^2} = 0 \Rightarrow N_{xx} \rightarrow 0,
\end{align}
but
\begin{align}
    f_y \rightarrow \frac{C - z}{B - y} \Rightarrow N_{yy} = \frac{1}{2\pi} \sum_{p_y,p_z = \pm 1} \arctan \frac{C + p_z z}{B + p_y y},   \label{eq:N_yy_inf}
\end{align}
and similarly
\begin{align}
    N_{zz} =\frac{1}{2\pi} \sum_{p_y,p_z = \pm 1} \arctan \frac{B + p_y y}{C + p_z z},   \label{eq:N_zz_inf}
\end{align}
so \cref{eq:N_yy_inf,eq:N_zz_inf} give the only non-zero components of $\mathrm{N}_\infty^\text{prism}$.


\section{Numerical validation \label{sec:numerical_validation}}

\begin{figure*}[htb]
    \centering
    \includegraphics[width=0.95\textwidth]{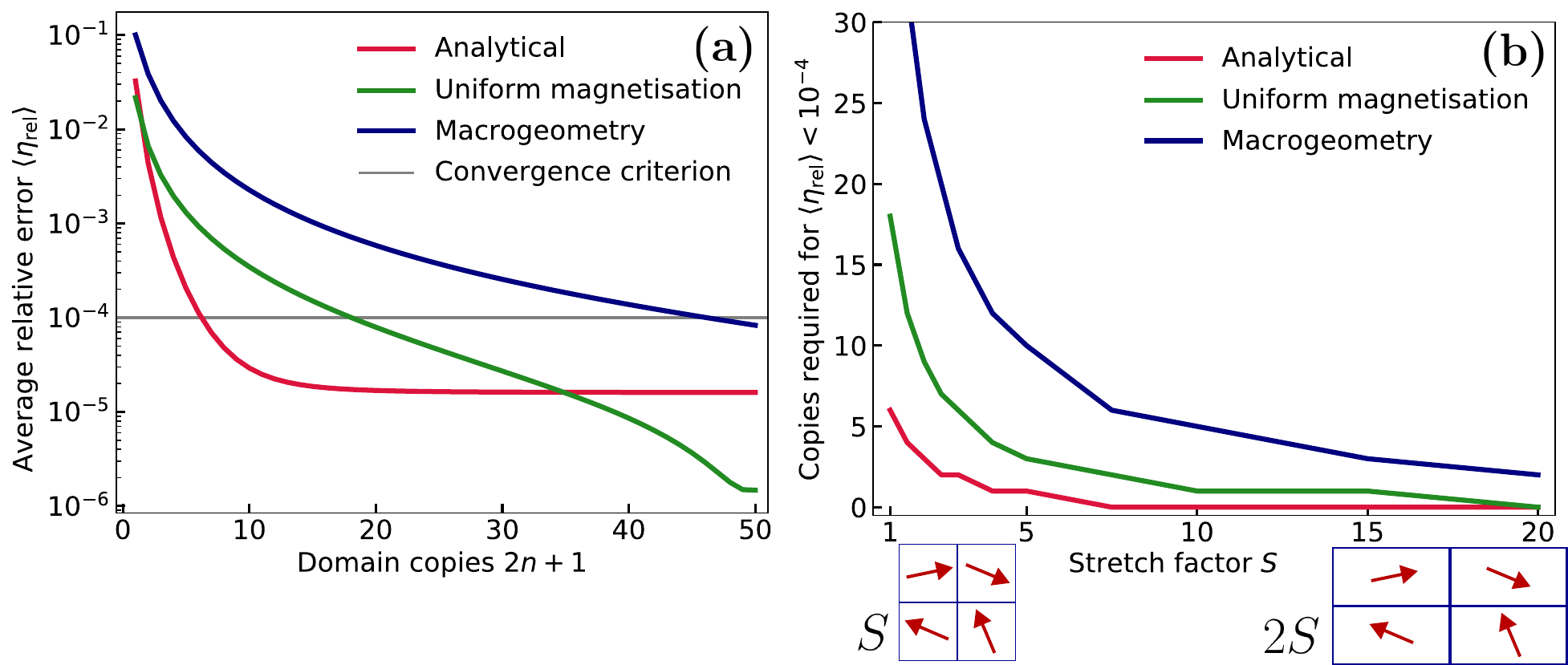}
    \caption{Convergence tests. \textbf{(a)} average relative error vs.\ the number of domain copies used in the macrogeometry. The blue line corresponds to $\mathrm{N}^\text{rem} = 0$ while the red and green lines use the analytical solutions of \cref{subsec:array_of_prisms,subsec:uniform_magnetisation} for $\mathrm{N}^\text{rem}$. The gray line is the chosen point of convergence, i.e.\ where the relative error is $10^{-4}$. \textbf{(b)} the number of domain copies where each method reaches convergence as function of the factor $S$ by which the entire system has been dilated along $x$. A doubling of the stretch factor for a 4-cell system is sketched below the plot.}
    \label{fig:convergence_tests}
\end{figure*}

To numerically verify the accuracy and utility of the analytical results from \cref{subsec:array_of_prisms,subsec:uniform_magnetisation} we consider a test system where the initial domain consists of 3116 variably sized cube cells forming a single cube of side length. The magnetisation at site $i$ is given by
\begin{align}
    \vb{M}_i = \frac{1}{\sqrt{3}} \mqty(1 \\ 1 \\ 1) + \vb{M}_i^\text{ran},
\end{align}
where $\vb{M}_i^\text{ran}$ has a uniformly distributed random direction and a magnitude drawn from a Gaussian distribution with standard deviation of 1. Having both a net magnetisation and random noise of comparable magnitude on a complex simulation domain ensures a non-trivial test.

As the reference field $\vb{H}_\text{ref}$ to which we will compare the derived solutions we use the pure macrogeometry method ($\mathrm{N}_i^\text{rem} = 0$) which is well-established\cite{vansteenkiste_design_2014,fangohr_new_2009}. To ensure convergence we set $n_\text{ref} = 125$, so 251 copies of the simulated domain are used to approximate $\mathrm{N}_i^\text{PBC}$. We define the relative error at position $\vb{r}$ by
\begin{align*}
    \eta_\text{rel}(\vb{r}) = \frac{\abs{\vb{H}(\vb{r}) - \vb{H}_\text{ref}(\vb{r})}}{\abs{\vb{H}_\text{ref}(\vb{r})}},
\end{align*}
and the average relative error $\expval{\eta_\text{rel}}$ by the average of $\eta_\text{rel}$ over a $20\times 20 \times 20$ grid of positions uniformly distributed across the initial domain. Specifically we sample between the points $(\pm A, \pm 0.999 B, \pm 0.999C)$ because the uniform magnetisation solution (\cref{eq:N_zz_inf,eq:N_yy_inf}) is ill-defined at the corner points where $y=\pm B$ and $z = \pm C$.

In \cref{fig:convergence_tests}a $\expval{\eta_\text{rel}}$ as function of macrogeometry copies is shown for the analytical solution from \cref{subsec:array_of_prisms}, the uniform magnetisation approximation of \cref{subsec:uniform_magnetisation} and the base macrogeometry method. We observe that while all three methods do converge, reaching relative errors less than $10^{-4}$, the convergence rate is significantly faster for the analytical solution. Besides validating our derivations, this justifies implementing the analytical solution in micromagnetics software for 1D PBCs and less specialised methods like the uniform magnetisation approach for 2D and 3D PBCs.

Interestingly the analytical solution plateaus at a finite error. Changing from double- to single precision digits in the numerical computation did not visibly change the error, suggesting machine precision is not a limitation, but increasing $n_\text{ref}$ from 75 to the current value of 125 decreased the error by about a factor 2. We conclude that because the macrogeometry method converges slowly, converging to less than $10^{-5}$ of the true value is impractical and the better agreement for uniform magnetisation is most likely coincidental.

We also tested the effect of dilating the system along $x$ by a factor $S$, i.e.\ stretching every constituent micromagnetic cell from cubes to prisms and shifting center positions accordingly. We use the same magnetisation between simulations, including the random components. In principle this dilation brings the calculation closer to the case in \cref{fig:thin_prism_array} where the analytical result is exact. In \cref{fig:convergence_tests}b we plot the number of copies, $n_\text{conv}$, need to obtain a relative error below $10^{-4}$, as function of stretch factor. We see that every method converges faster, because dilation moves distant domain copies further away hence reducing their relative contribution, but the fastest converging method remains the analytical solution in all cases with about an order of magnitude reduction in $n_\text{conv}$ over the base macrogeometry method.

\section{Discussion and conclusion \label{sec:conclusion}}

In this paper we presented three analytical solutions for the magnetostatic field from a system of uniformly magnetised cells that is periodic along one dimension. Two of the solutions are exact in special, limiting cases, i.e.\ for the on-axis field in an array of point dipoles and an array of infinitesimally thin rectangular prisms.

Besides the intrinsic value of exact solutions in magnetostatics, which are rare in periodic systems, we demonstrated the applicability of our results to micromagnetic simulations. When supplementing the macrogeometry method for nearby regions with the analytical prism solution for more distant parts of the system, the number of domain copies in the macrogeometry required for convergence was decreased by nearly an order of magnitude. The prism solution is restricted to 1D PBCs, where convergence is relatively easy as the computational cost scales linearly with the characteristic length of the macrogeometry. Nevertheless, our study demonstrates the usefulness of analytical solutions for high-precision calculations or when the simulated domain is particularly complex. With 2D or 3D PBCs the number of domain copies grows non-linearly with the macrogeometry scale, so for complex systems, the uniform magnetisation approximation or equivalent can be crucial for converging the demagnetisation tensors.

\section*{Acknowledgement}
This work was supported by the Villum Foundation Synergy project number 50091 entitled "Physics-aware machine learning" and the Carlsberg Foundation Semper
Ardens Advance project CF24-0920 entitled ``Novel magnets through interdisiplinarity and nanocomposites''.

\section*{Data statement}
All data presented in this work are available from Ref. 


\bibliographystyle{abbrv}
\bibliography{references}

\onecolumngrid

\appendix



\end{document}